\documentclass[prc,showpacs,twocolumn]{revtex4-2}
\usepackage{amssymb}
\usepackage{natbib}
\usepackage{graphicx}
\usepackage{dcolumn}
\usepackage{bm}
\usepackage{multirow}
\usepackage{amsmath}
\usepackage{hyperref}
\usepackage{subfigure}
\usepackage{amssymb}

\begin{document}
        
        \title{Examining potential energy surface through Chebyshev shape parametrization}
        
        \author{K. Jyothish}
        \email{jyothishk003@gmail.com, jyothishk003@cusat.ac.in}
        \author{M. S. Suryan Sivadas}
        \email{ms3.msssivadas@cusat.ac.in}
        \author{A. K. Rhine Kumar}
        \email{rhinekumar@cusat.ac.in}
        \affiliation{Department of Physics, Cochin University of Science and Technology, Kochi--682022, India.}
        \date{\today}
        
\begin{abstract}

The present study introduce a novel approach, the Chebyshev shape parametrization, to describe the geometric configurations of atomic nuclei, with a particular emphasis on fission dynamics. In this framework, the nuclear surface is represented by a profile function expanded in a Chebyshev polynomial series, with deformation parameters derived analytically under volume conservation and centre-of-mass constraints. The proposed parametrization is shown to be universal robust, and we establish transformation equations that connect it to other widely used shape parametrizations.
In the macroscopic approach, the potential energy surface (PES) is computed using the Lublin–Strasbourg Drop (LSD) model, incorporating deformation-dependent energy coefficients expressed in terms of Chebyshev parameters. This enables a detailed investigation of the structural evolution and fission pathways of the nucleus $^{227}$Pa across various deformations, depicting the influence of shape parameters on elongation, asymmetry, and neck formation.
Complementing this, microscopic analysis is carried out by calculating single-particle energy levels through diagonalization of the Yukawa-folded mean-field Hamiltonian in a deformed harmonic oscillator basis. The nuclear shape parameters are provided by the Chebyshev parametrization, allowing us to examine shell structure effects at specific deformations. Together, these macroscopic and microscopic studies provide comprehensive insight into the nuclear energy landscape and shape evolution during fission.

        
\end{abstract}

\maketitle
\section{\label{sec:level1}Introduction\protect }

In recent years, significant theoretical developments have been made in the dynamic reshaping of atomic nucleus during the fission process \cite{Theory_of_nuclear_fission,Bohr, krappe_pomorski_2012}. However, a comprehensive theoretical framework that explains the complexities of fission reactions continues to be challenging. The structural and dynamic evaluation of fission requires three essential components: nuclear shape parameterization, the total potential energy of the nucleus in terms of its deformation, and the differential equation describing its time evolution  \cite{Mazurek}. 
The primary objective is to select an appropriate shape parameterization or profile function to determine the shape of the nucleus \cite{optimal,MYERS}. This function must converge rapidly with a minimal number of parameters to ensure simplicity and accuracy. It should be sufficiently flexible to represent a broad range of nuclear shapes, from spherical to highly distorted configurations, including dumbbell structures. Furthermore, the function must be compatible with current theoretical models and ensure computational efficiency. Various shape parametrization methods have been incorporated into theoretical models to explore the potential energy landscape of atomic nuclei \cite{MOLLER,geometrical,Ivanyuk}.
Among these methods include Bohr-Mottelson shape parametrization $(\beta,\gamma)$  \cite{spherical}, Cassini ovals  \cite{Cassini1,Cassini2}, quadratic surfaces of revolution (QSR) \cite{QSR1,QSR2}, Funny-Hills (F-H) shapes \cite{Funny, Mazurek2,Mazurek3}, modified Funny-Hills \cite{Modified_funny}, Trentalange-Koonin-Sierk (TKS) shape parametrization \cite{TKS}, and Fourier shape parametrization \cite{Fourier1}. Still the scientific communities are exploring new parametrization methods to investigate all the changes during the fission.  The present work proposes a new, rapidly converging shape parametrization, the Chebyshev shape parametrization, to describe all the possible geometric configurations of the atomic nucleus.\\
The second key objective is to evaluate the total potential energy of the nucleus. Nuclear theoretical frameworks to unveil the potential energy landscapes are categorized into different paradigms. The first encompasses the microscopic approach, including the relativistic mean field model \cite{MFT1, RevModPhys.75.121}, Hartree-Fock-Bogoliubov with Skyrme \cite{hfbs1, hfbs2} or Gony forces \cite{hfbg}, quantum field theory \cite{RevModPhys.75.121}, and density functional theories \cite{MENG2006470}. The second category includes phenomenological macroscopic models developed following George Gamow's liquid drop model (LDM) \cite{gamow1930,Weiz,Bethe}, such as the droplet model (DM), the finite-range droplet model (FRDM) \cite{MOLLER3}, and the Lublin-Strasbourg drop (LSD) model  \cite{pomorski-curv}. LSD model relatively modernized LDM  by incorporating the curvature and congruence terms expressed through Leptodermous expansion. By the 1970s, it became clear that the LDM was inadequate for replicating measured fission barrier heights. This realization led to the development of extensions to LDM with microscopic contributions as proposed by Strutinsky \cite{STRUTINSKY1,STRUTINSKY2}. The macroscopic-microscopic methods calculates the potential energy as the sum of shape-dependent macroscopic and microscopic terms. These shape-dependent terms make them computationally efficient, allowing for rapid predictions within a multidimensional deformation space across the nuclear landscape \cite{moller4, MYERS4, MYERS2, pomorski-curv, MYERS1976411}. The macroscopic-microscopic method is notable for its effectiveness in explaining various aspects of atomic nuclei, including shape coexistence \cite{moller4,Shape_coexistence}, hyperdeformation \cite{Dudek2003}, stability of super heavy nuclei \cite{Pomorski2022}, and giant dipole resonances \cite{rhine}. However, as temperature increases, the microscopic contribution diminishes, leaving the macroscopic contribution dominant \cite{MYERS}.\\
In the present scenario exploring a new shape parametrization to define the nuclear geometry and study the resulting changes in the potential energy has relevance. We define the nuclear geometry using Chebyshev shape parametrization and the deformation induced changes in the potential energy are explained by exerting the LSD model.\\
Sec. \ref{sec2}  presents the theoretical framework, divided into three subsections. Sec. \ref{sec2a} details the proposed profile function for nuclear geometry and the essential quantities for formulating various nuclear shapes. Sec. \ref{sec2b} establishes the relationship between Chebyshev shape parametrization and other existing parametrization techniques. Sec. \ref{sec2c} discusses the model used to determine the total potential energy. The results are presented in Sec. \ref{sec3}, where a detailed study of the potential energy surface as a function of different deformation parameters is conducted. The conclusions are detailed in Sec. \ref{sec4}.
 \section{Theoretical framework}
 \label{sec2}
 \subsection{Chebyshev shape parametrization}
 \label{sec2a}

Chebyshev polynomials play a significant role in numerical analysis and approximation theories, providing advantages over the Power, Legendre, and Fourier series. The orthogonality of Chebyshev polynomials minimizes round-off error accumulation, addressing the numerical instability often associated with the power series \cite{mason_handscomb_2002}. 
With nodes clustered at the boundaries, Chebyshev interpolation reduces Runge's phenomenon and offers better accuracy than evenly spaced nodes in power series. In contrast, Legendre nodes, denser at the centre, tend to amplify it \cite{boyd2001chebyshev}. Unlike the Fourier series, which is optimized for periodic functions, Chebyshev polynomials are naturally suited for bounded and non-periodic domains [-1,1], making them ideal for modelling irregular geometrical shapes and solving boundary-value problems . Moreover, they are less prone to the Gibbs phenomenon, offering improved accuracy for sharp transitions and boundary-layer effects \cite{Gibbs}.  The exponential convergence of the Chebyshev series for analytic functions within the ellipse of convergence surpasses that of the Legendre and the power series \cite{Trefethen2013}. 
They are particularly valued for their minimax property, which minimizes the maximum error in polynomial approximations \cite{chebyshev}.
Efficient evaluation algorithms, such as the Clenshaw algorithm \cite{Clenshaw_1957}, make the computation of Chebyshev polynomials straightforward and fast. This computational efficiency is advantageous in real-time applications and in large-scale numerical simulations.
 
 We developed a shape parametrization method using Chebyshev polynomials of the first kind ($T_n(x)$), addressing the limitations of existing parametrization methods. They excel in representing non-periodic phenomena within confined spaces, making them particularly useful for understanding the potential energy landscape of the atomic nuclei. 
 Nuclear shape parametrization as a profile function using Chebyshev series expansion in a cylindrical coordinate ($r,\phi,z$) system is written as,
 \begin{equation}
        \label{1}
        \frac{{\rho}^2_s(u)}{R_0^{2}}=\sum_{n=0}^{\infty}a_n T_n\left(u\right) ,
 \end{equation}
 
 \begin{equation*}
        T_n(u)=\cos[n\cos ^{-1}(u)] \quad \quad \text{if} \quad -1\le u\le 1.
 \end{equation*}
 \begin{figure}
        \centering
        \includegraphics[width=\linewidth]{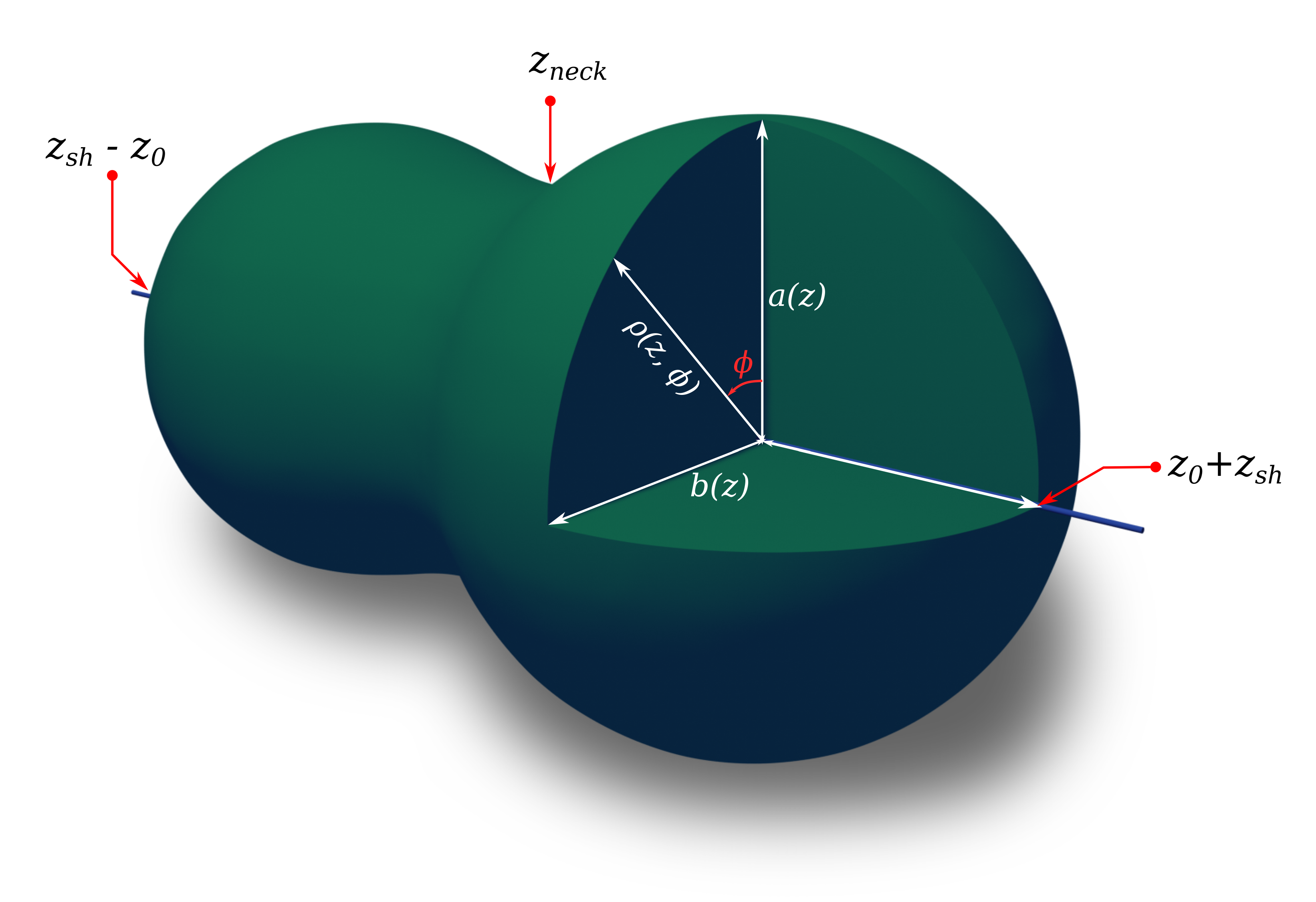}
        \caption{Schematic representation of the geometry of nucleus in cylindrical coordinates using the Chebyshev profile function.}
        \label{fig:1}
 \end{figure}
 ${\rho}_s(u)$ represents the perpendicular distance from a surface point to the symmetry axis ($z$ axis), where $u=\frac{z-z_{sh}}{z_0}$, $R_0$ denotes the radius of a spherical shape with the same volume as the nuclear shape being considered. $a_n$ is the Chebyshev series coefficient, which is considered the deformation parameter for explaining different shapes of the nucleus. Along the symmetry axis, the shape extends by a length of $2z_0$, with its left and right ends positioned at $z_{\text{min}}=z_{sh}-z_{0}$ and $z_{\text{max}} = z_{sh} + z_{0}$, respectively. $z_{sh}$ is the shift coordinate, which will be discussed in the section \ref{sec2a2}.  The boundary condition of the profile function can be written as, ${\rho}^2_s(z_{\text{min}})={\rho}^2_s(z_{\text{max}})=0$,
 which gives the following relation, 
 \begin{equation}
        \label{2}
        \begin{aligned}
                \sum_{n=0}^\infty a_n T_n(-1)=\sum_{n=0}^\infty(-1)^n a_n =0, \\
                \sum_{n=0}^\infty a_n T_n(1)=\sum_{n=0}^\infty a_n =0. 
        \end{aligned}
 \end{equation}
 $a_0$ and $a_1$ are estimated from the above conditions,
 \begin{equation}
        \label{3}
        a_0 =-\sum_{n=2,4,..}^\infty a_n,\ \ a_1 =-\sum_{n=3,5,..}^\infty a_n.
 \end{equation}
 \subsubsection{Volume conservation}
 \label{sec2a1}
 The nucleus is treated as an incompressible liquid drop with a constant volume. Thus, the volume equivalence between a sphere and a deformed nucleus can be stated as,
 \begin{equation}
        \label{4}
        \begin{split}
                V &= \frac{4\pi}{3}R_0^3 = \int_0^{2\pi} d\phi \int_{z_{\min}}^{z_{\max}} dz \int_0^{{\rho}_s(z)} \rho \, d\rho ,\\
                &= \pi \int_{z_{\min}}^{z_{\max}}{\rho}_s^2(z) \, dz .
        \end{split}
 \end{equation}
 The length of the nucleus in terms of elongation $(c)$ is defined as $z_{0}=cR_{0}$, which can be obtained by simplifying Eq.~\ref{4} and relationship between $c$ and all even coefficients, $a_{2n}$ is formulated as,
 \begin{equation}
        \label{5}
        \sum_{n=0}^{\infty}\frac{a_{2n}}{1-4n^2}=\frac{2}{3c} .  
 \end{equation}
 Eq.~\ref{5} is analogous to the volume conservation relation in the Funny-Hills (F-H) parameterization and the Fourier shape parametrization \cite{NADTOCHY, BARTEL, Pomorski_2017, Nerlo-Pomorska_2015}. This volume conservation equation can find $a_2$ and its higher order as a function of elongation $(c)$. The nucleus adopt a spherical shape when $c$ equals one, however, if $c$ decreases or increases from this value, the nucleus exhibits either an oblate or prolate shape, respectively. 
 Considering the nucleus in a spheroidal condition by limiting Eq.~\ref{5} to $n=1$, we can obtain the relation $c=-1/{2a_2}$. The spherical, prolate and oblate nuclear shapes can be obtained by the conditions $a_2=-\frac{1}{2}$, $-\frac{1}{2}<a_2<0$  and $a_2<-\frac{1}{2}$ respectively, where $a_2$ is the sole shape-determining parameter.
 For higher values of $n$, additional shape-determining parameters emerge, potentially leading to the breakdown of the simple spheroidal condition and the formation of more complex structures, such as a dumbbell shape.
 
 \subsubsection{Centre of mass formalism}
 \label{sec2a2}
 Given the inherent asymmetry in the shape of the system, fixing the system's centre of mass becomes necessary. The general equation with respect to the $z$-axis can be described mathematically as \cite{Fourier1},
 \begin{equation}
        \label{6}
        z_{cm}=\frac{\int_v z\rho d^3r}{\int_v \rho d^3r}=\frac{\pi\int_{z_{\text{min}}}^{z_{\text{max}}}\rho_s^2(z)z dz}{\pi\int_{z_{\text{min}}}^{z_{\text{max}}}\rho_s^2(z)dz}=0.
 \end{equation}
 To address the geometry of the left-right asymmetry in the nucleus during the fission process, a new term was introduced, named as shift coordinate ($z_{sh}$), which ensures that the center of mass of the nucleus is positioned at the origin. This quantity is zero when all the odd deformation parameters $a_n$ are equal to zero. This shift coordinate is derived from the Eq.~\ref{6} and expressed as,
 \begin{equation}
        \label{7}
        z_{sh}=\frac{3cz_0}{2}\sum_{n}\frac{a_{2n+1}}{{(2n+1)^2}-4}.
 \end{equation}
 \subsubsection{Distance between nascent fragments}
 \label{sec2a3}
 Determining the distance between the centre of mass of the nascent left-right fragments ($R_{12}$) is essential for fission studies, which can be obtained as, 
 \begin{equation}
        \label{8}
        R_{12} = \frac{\pi\int_{z_{\text{neck}}}^{{z_{\text{max}}}}\rho_s^2 (z)zdz}{\pi\int_{z_{\text{neck}}}^{{z_{\text{max}}}}\rho_s^2 (z)dz}- \frac{\pi\int_{z_{\text{min}}}^{{z_{\text{neck}}}}\rho_s^2 (z)zdz}{\pi\int_{z_{\text{min}}}^{{z_{\text{neck}}}}\rho_s^2 (z)dz}.
 \end{equation}
 Here, $z_{\text{neck}}$ represents the position at which the fission fragments are formed. At this point, the function $\rho_s$ reaches its minimum, and the nucleus approaches the scission point, corresponding to $\frac{d\rho_s}{dz}=0$. To facilitate the analysis, a new coordinate $u_{\text{neck}}$ can be introduced, which is defined as    $u_{\text{neck}}=\frac{(z_{\text{neck}}-z_{sh})}{z_0}$. 
 From the minimization condition, a relation can be obtained as,
 \begin{equation}
        \label{9}
        R_{12}=\frac{\sum_{m=0}^\infty (-1)^m a_{2m+1} (2m+1)}{\sum_{m=0}^\infty (-1)^m a_{2m} 4m^2},
 \end{equation}
 by retaining terms only up to $a_4$ in the summation, the neck thickness can be approximated as,

 \begin{equation}
        \label{10}
        R_{12}=\frac{a_3}{a_2-4a_4}.
 \end{equation}

 \subsubsection{Realistic profile function: Non-axial case}
 \label{sec2a4}
 Since the nucleus' deformation might not be strictly axial, we have included a non-axial parameter to better capture its true structure. Consider the equation of ellipsoid, which depends on the deformation of nucleus \cite{DOBROWOLSKI},
 \begin{equation}\label{11}
        \frac{x^2}{a_x^2}+\frac{y^2}{a_y^2}=1,
 \end{equation}
 where $a_x$ and $a_y$ are half-axis parameters. The equation of ellipsoid in polar coordinates is written as,  
 \begin{equation}\label{12}
        \rho^2(z,\phi)=\frac{a_x^2a_y^2}{a_y^2\cos^2(\phi)+a_x^2\sin^2(\phi)},
 \end{equation}
 The non-axiality parameter $\eta$ is the relative difference of the half axis of the cross-section perpendicular to the symmetry axis and is given by \cite{Pomorski2022},
 \begin{equation}\label{13}
        \eta=\frac{a_y-a_x}{a_y+a_x}.
 \end{equation}
 It describes the extent of non-axiality occurs due to the deformation of nucleus. For an axially symmetric shape, $\eta$ becomes zero. Here, we consider the volume of the deformed nucleus to be the same as that of the spherical nucleus. Hence, the volume conservation equation becomes, 
 \begin{equation}\label{14}
        \pi\rho^2=\pi a_x a_y. 
 \end{equation}
 For the pure ellipsoidal shape, the half axis can be described in terms of deformation parameters  $c$ and $\eta$ as follows,
 \begin{equation}
        \label{15}
        a_x=\frac{R_0}{\sqrt{c}}\left({\frac{1-\eta}{1+\eta}}\right)^{1/2}, 
        a_y =\frac{R_0}{\sqrt{c}}\left({\frac{1+\eta}{1-\eta}}\right)^{1/2},  
        a_z = R_0c .
 \end{equation}
 A non-axial nuclear shape can be derived by incorporating the $\eta$ parameter into the axial-profile function (${\rho}_s(z,\phi)$).  This approach leads to a more realistic representation of the nuclear profile function, which is expressed as,
 \begin{equation}
        \label{16}
        \rho ^2(z,\phi)={\rho}_s^2(z)\frac{1-\eta^2}{1+\eta^2+2\eta \cos(2\phi)}. 
 \end{equation}
 A schematic diagram depicting an elongated, dumbbell shaped, axially symmetric nuclear surface along the $z$-axis is illustrated in Fig.~\ref{fig:1}.

 \subsubsection{Convergence properties}
 \label{sec2a5}
 \begin{figure}
        \centering
        \includegraphics[width=1\linewidth]{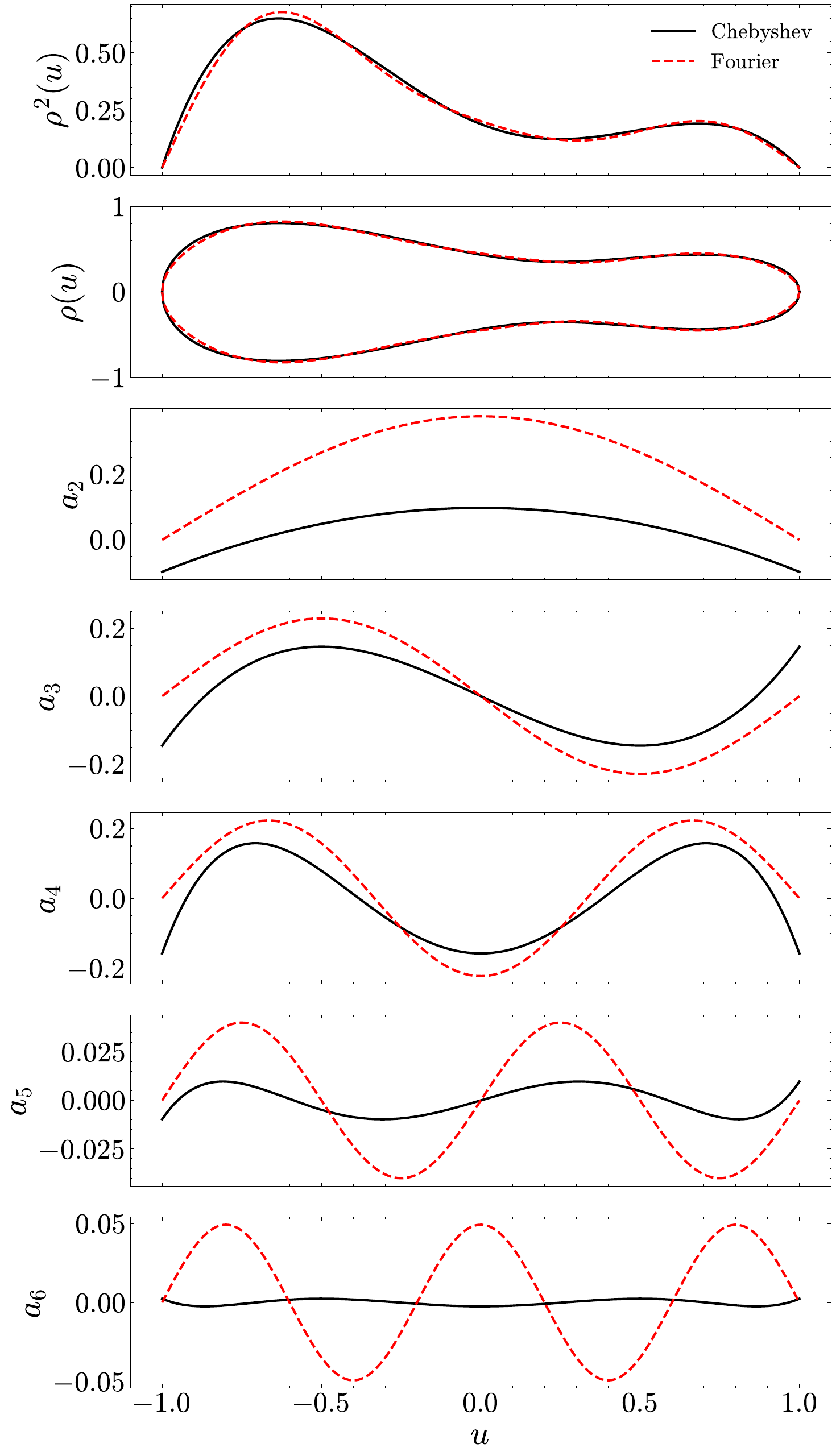}
        \caption{Comparison of the convergence between  Chebyshev and Fourier shape parametrizations for a specific nuclear shape, depicted through subplots of the profile functions $\rho^2(u)$ and $\rho(u)$, alongside contributions from different orders of expansion ($a_2$ to $a_6$) as functions of $u$.
        }
        \label{fig:1a}
 \end{figure}
 
 \begin{figure*}
        \centering
        \includegraphics[width=0.99\linewidth]{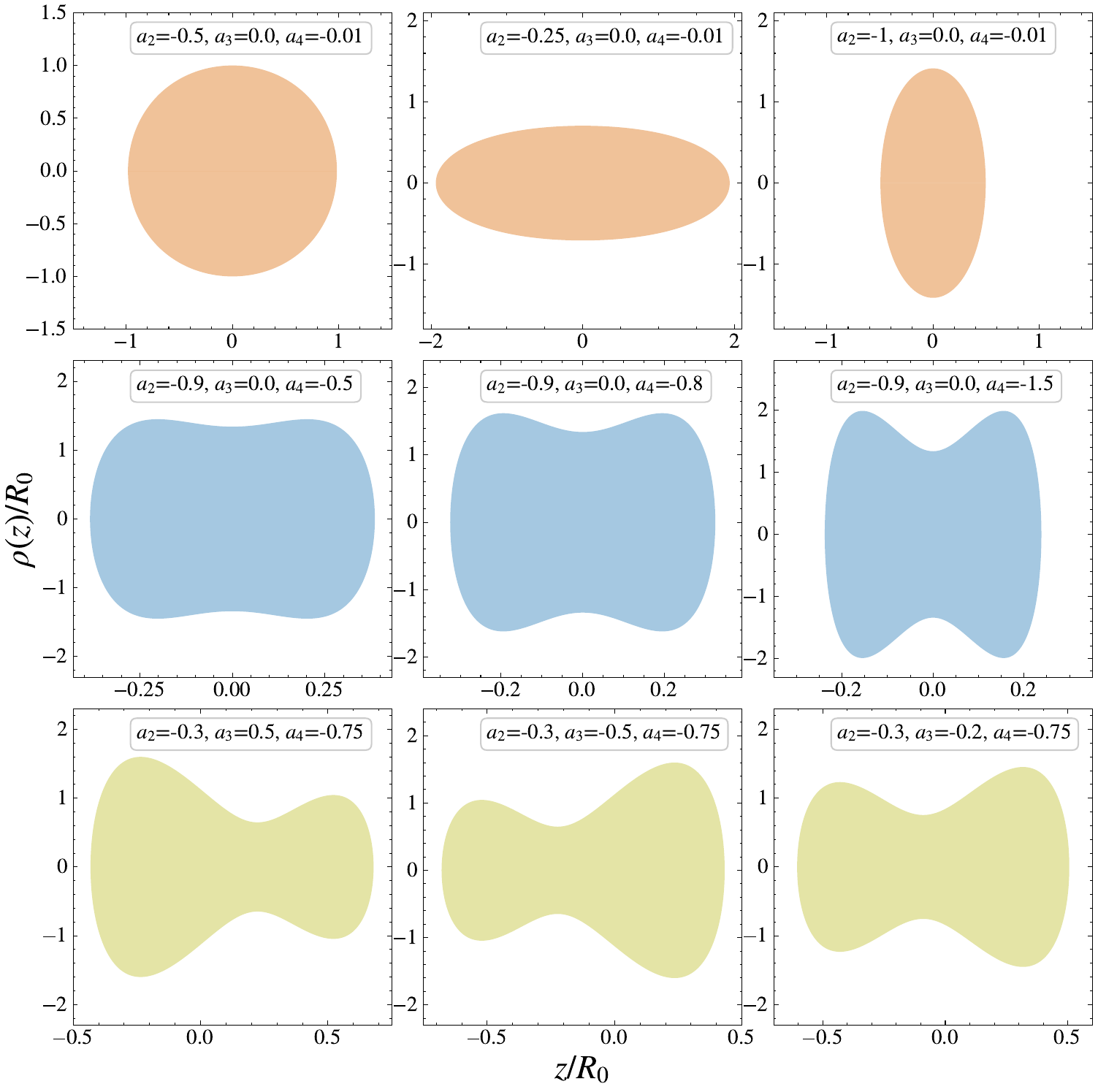}
        \caption{Possible nuclear shapes using Chebyshev shape parametrization with different deformation parameters.}
        \label{fig:2}
 \end{figure*}

 A key advantage of using the Chebyshev parametrization, as shown in Eq. \ref{1}, is its flexibility in precision. This approach can be extended to any desired order, allowing for a detailed study of its convergence properties.
 
 Fig. \ref{fig:1a} compares the convergence of Chebyshev and Fourier shape parametrizations for a specific nuclear shape through the profile function and its coefficients. The figure consists of seven subplots: the nuclear profile function $\rho^2(u)$, $\rho(u)$, and the contributions from different orders of expansion to the shape function, specifically, $a_2 ,a_3, a_4 ,a_5$ and $a_6$ as functions of $u$. Chebyshev coefficients exhibit faster convergence across orders than Fourier coefficients, indicating greater efficiency in accurately representing nuclear surface deformations. The figure also shows that the contributions of $a_5$ and $a_6$ are almost negligible in Chebyshev shape parametrization compared to Fourier shape parametrization, further supporting its efficiency.
 
The $a_2$ component has a value of $0.2$, while $a_6$ drops to approximately $0.01$, over an order of magnitude lower. Similarly, $a_5$, at $0.02$, is also more than an order of magnitude smaller than  $a_3$ which is approximately $0.3$. This pattern illustrates how higher order terms contribute minimally, enabling precise control over shape details with only a few leading terms. The lowest-order term, $a_2$ primarily controls the elongation of shape, while the next term, $a_3$ introduces left-right asymmetry $a_4$ defines the neck formation  and $a_5$ adds only minor distortions in left-right asymmetry. The parameter $a_6$ further refines the neck with minimal adjustment. Generally, higher-order Chebyshev coefficients (both even and odd) have a small or negligible impact on the shape. To determine the potential energy surface, we utilized three Chebyshev parameters $a_2$, $a_3$ and $a_4$  along with an additional non-axiality parameter, $\eta$. 
 
 Using Eq.~\ref{1}, a series of nuclear shapes characterized by deformation parameters $a_2$, $a_3$, and $a_4$ have been plotted and are shown in Fig.~\ref{fig:1a}. In the first row, $a_3$ and $a_4$ are fixed, and $a_2$ is varied, generating spherical, prolate, and oblate shapes. It demonstrates that $a_2$ characterizes the elongation of the nucleus.  In the second row,  $a_2$ and $a_3$ are fixed and $a_4$ is varied. By fixing $a_3=0$, we got a symmetric dumbbell shape. By gradually increasing the negative value of $a_4$, the relative neck thickness of the shapes increases, indicating that $a_4$ controls the neck formation. In the third row, $a_2$ and $a_4$ are fixed, while $a_3$ is varied. This results in left and right asymmetric shapes for $a_3 =0.5$ and $-0.5$, respectively, with a moderately asymmetric shape plotted for $a_3 =-0.2$.  Therefore,  the odd parameters of the Chebyshev polynomials characterize the asymmetric shape of the nucleus, and the even parameters control the relative elongation and neck formation of the nucleus. Hence, the Chebyshev shape parametrization effectively captures the diverse shapes during fission.
 \subsection{Relation with other shape parametrization}
 \label{sec2b}
 \subsubsection{Bohr-Mottelson shape parametrization $(\beta,\gamma)$}
 \begin{figure}
        \centering
        \includegraphics[width=1\linewidth]{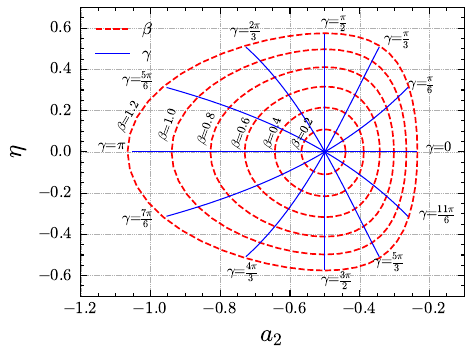}
        \caption{The relationship between traditional Bohr-Mottelson shape parameters $(\beta,\gamma)$ and Chebyshev parameters $(a_2,\eta)$. }
        \label{fig:3}
 \end{figure}
 Bohr-Mottelson shape parametrization $(\beta,\gamma)$, is suitable for minor to moderate deformations, and becomes complex for highly deformed structures with substantial number of terms up to $\beta_{14}$ to accurately represent nuclear shapes and their liquid-drop energy during the fission process \cite{schunck,PhysRevC.87.044308, nix}. Furthermore, the function does not converge rapidly when a nucleus is highly elongated, particularly between the saddle and scission points \cite{Modified_funny}.
 The relationship between generally used ($\beta,\gamma$) and the ($\eta, c$)  can be derived from the half-axis formula of the ellipsoid \cite{elipseaxis}. The half-axis formula in terms $\beta$ and $\gamma$ is defined as \cite{Pomorski2022},
 
 \begin{align}
        R_x &= R_0\exp{\left(\sqrt{\frac{5}{4\pi}}\beta\cos{\left(\gamma-2\pi/3\right)}\right)} 
        = \frac{R_0}{\sqrt{c}}\left({\frac{1-\eta}{1+\eta}}\right)^{1/2} \nonumber\\
        R_y &= R_0\exp{\left(\sqrt{\frac{5}{4\pi}}\beta\cos{\left(\gamma-4\pi/3\right)}\right)} 
        = \frac{R_0}{\sqrt{c}}\left({\frac{1+\eta}{1-\eta}}\right)^{1/2} \nonumber\\
        R_z &= R_0\exp{\left(\sqrt{\frac{5}{4\pi}}\beta \cos{\gamma}\right)} 
        = R_0c \nonumber
 \end{align}
 Using the above equations, we can establish the relationship between ($\beta$, $\gamma$) and ($\eta$, $a_2$).
 \begin{equation}
        \label{17}
        \beta=\left\{\frac{4\pi}{5}\left[\frac{1}{3}\log^2\left(\frac{1+\eta}{1-\eta}\right) + \log^2\left(-\frac{1}{2a_2}\right)\right]\right\}^{1/2}
 \end{equation}
 \begin{equation}
        \label{18}
        \gamma=\tan^{-1} \left[\frac{1}{\sqrt{3}}\frac{\log\left(\frac{1+\eta}{1-\eta}\right)}{\log\left(-\frac{1}{2a_2}\right)}\right]
 \end{equation}
 Fig. \ref{fig:3} illustrates the relationship between $(\beta,\gamma)$ and $(a_2,\eta)$, obtained from Eqs.~\ref{17} and \ref{18} respectively. This relationship holds only under spheroidal deformation, and breaks down if the system adopts a dumbbell shape. The figure demonstrates a $60^o$ symmetry, indicating that the exact shape repeats every $60^o$. 
 For example, consider an axially symmetric prolate shape at deformation parameters ($\beta=0.1, \gamma=0^o$) with  $(a_2\approx-0.47, \eta=0)$. This shape is identical to the one at $(\beta=0.1, \gamma=120^0)$ with  $(a_2\approx-0.52, \eta\approx0.06)$. Similarly, an axially symmetric oblate shape at $(\beta=0.6,\gamma=180^o)$ with $(a_2\approx-0.73, \eta=0)$ is equivalent to the shape at $(\beta=0.6, \gamma=60^o)$ with  $(a_2\approx-0.42, \eta\approx0.28)$. These equivalences occur for spheroidal shapes when higher-order Chebyshev shape parameters ($a_n$ for $n>$2) are neglected. 
 It is observed that a larger value of $\beta$ can't be attributed to the elongation of shape, and a non-axial parameter, $\eta$, will have a finite value as illustrated in Fig.~\ref{fig:3}. This observation indicates the $(a_2,\eta)$  deformation space is more appropriate for exploring the triaxial shapes than the conventional ($\beta, \gamma$) space. Hence, Chebyshev shape parametrization offers a more comprehensive framework for describing the shape of nucleus.  
 \subsubsection{Funny-Hills shape parametrization}
 F-H shape parametrization introduces additional solutions between the boundary conditions for certain combinations of deformation parameters, which leads to negative values of $\rho_s^2(z)$. This illogical and unphysical outcome necessitates the modifications to this parametrization. Although the modified F-H parametrization provides valuable insights into the fission barrier, it often struggles to accurately predict the energies of fissioning nucleus, with limited improvements achievable by adjusting the deformation parameters \cite{mody-funny1}.
 
 The F-H shape parametrization for $B\ge0$ can be regarded as a special case of the Chebyshev shape parametrization. The relationship between these two parametrizations can be established by comparing the coefficients of their respective variables. 
 \begin{equation}
        \label{19}
        \begin{split}
                a_2&=-\frac{c^2A}{2}\\
                a_3&=-\frac{c^2\alpha}{4}\\
                a_4&=-\frac{c^2B}{8}
        \end{split}
 \end{equation}
 Where,         $A, B,$ and $\alpha $ are the F-H deformation parameters and $c$ is the common elongation parameter.
 \begin{figure*}
        \subfigure[]{
                \includegraphics[width=0.49\linewidth]{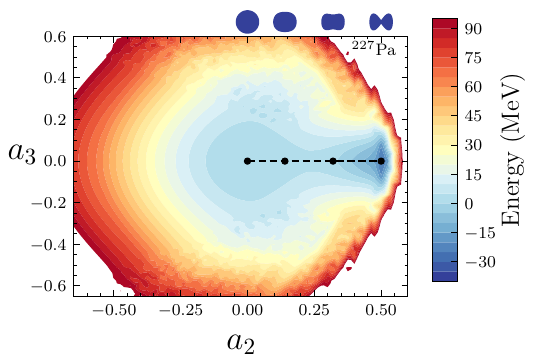}}
        \subfigure[]{
                \includegraphics[width=0.49\linewidth]{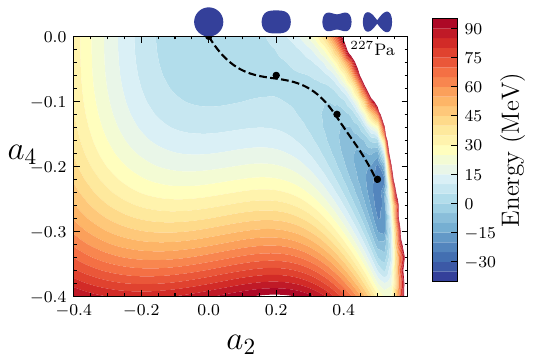}}
        \caption{Cross sections of PES of $^{227}$Pa at $L=0$ in the two-dimensional deformation subspace defined by (a) ($a_2, a_3$), and (b) $(a_2, a_4)$, remaining parameters are fixed at values corresponding to the minimum potential energy of the fissioning nucleus.}
        \label{Fig:4}
 \end{figure*}
 
 \subsubsection{TKS shape parametrization}
 The precision of TKS parametrization based on the Legendre polynomial approach can be improved by the higher-order expansions. However, it is constrained by the requirement of $\rho^2_s(z)\ge0$, which limits the number of deformation parameters.
 
 Similarly, a transformation relation between the Chebyshev and TKS shape variables ($\alpha_2 ,\alpha_3, \alpha_4$) can be formulated and approximated up to the fourth degree as given below,
 \begin{equation}
        \label{19-a}
        \begin{split}
                a_2&= \frac{3}{4}\alpha_2 -\frac{5}{16}\alpha_4\\
                a_3&= \frac{5}{8}\alpha_3\\
                a_4&= \frac{105}{192}\alpha_4.
        \end{split}
 \end{equation}
 
 \subsubsection{Fourier shape parametrization}
 The Fourier shape parametrization has recently emerged as a promising alternative, facilitating rapid convergence properties that provide crucial information about the fission process near the scission point. It offers detailed descriptions of highly elongated and necked nuclear shapes utilizing only a few deformation parameters. \cite{Kostryukov_2021,liu, PhysRevC.108.024605,PhysRevC.107.054616}.
 The evaluation of the relation between Coefficients of Fourier shape parametrization and Chebyshev shape parametrization is as follows,
 \begin{equation}
        \begin{split}
                a_0 &= \sum_{n=1}^{\infty}f_{2n}J_0\left(\frac{(2n-1)\pi}{2}\right)\\
                a_1 &= 2 \sum_{n=1}^{\infty}f_{2n+1}J_1\left( \frac{2n\pi}{2}\right)\\
                a_{2m}&= \sum_{m=1}^{\infty}\left[2(-1)^m\sum_{n=1}^{\infty}f_{2n} J_{2m}\left(\frac{(2n-1)\pi}{2}\right) \right]\\
                a_{2m+1} &= \sum_{m=1}^{\infty}\left[2(-1)^m \sum_{n=1}^{\infty}f_{2n+1}J_{2m+1}\left( \frac{2n\pi}{2}\right)  \right]
        \end{split}
 \end{equation}
 where $J_n$ is the $n^{th}$-order Bessel function of first kind, and $f_n$ represents the Fourier shape parameters. 
 Therefore, we successfully demonstrated that the Chebyshev shape parametrization is more universal, consistent, and robust since it can establish a significant relationship between all the  existing shape parametrizations.
 \subsection{Potential energy calculation}
 \label{sec2c}
 Following the determination of the nuclear geometry, the next phase entails a comprehensive characterization of the system's potential energy throughout the fission process. The energy of the nucleus is computed via two complementary frameworks, the macroscopic and microscopic approaches. Each employs Chebyshev shape parametrization to compute the nuclear potential energy for different deformation parameters.
 
 \subsubsection{Macroscopic approach}

 \begin{figure*}
        
        \subfigure[]{
                \includegraphics[width=0.47\linewidth]{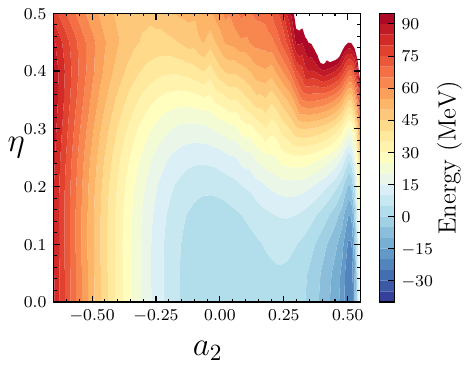}}
        \subfigure[]{
                \includegraphics[width=0.5\linewidth]{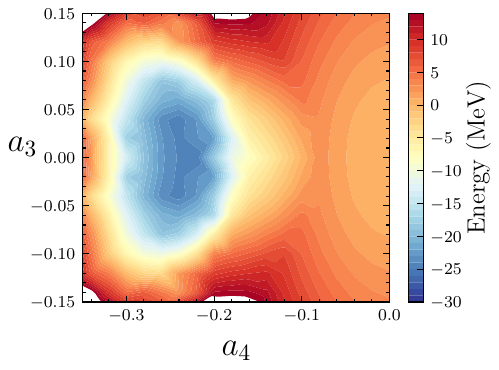}}
        \caption{Similar to Fig.~\ref{Fig:4}, but for the deformation subspace defined by (a) ($a_2, \eta$), and (b) $(a_4, a_3)$.}
        \label{Fig:5}
 \end{figure*}
To calculate the macroscopic energy, the Lublin–Strasbourg Drop (LSD) model is employed among the various liquid drop formulations. Unlike the traditional Liquid Drop Model (LDM), the LSD model includes additional energy contributions such as curvature and congruence terms, allowing for a more refined evaluation of the macroscopic potential energy of the system. The potential energy terms can be expressed as a leptodermous expansion with respect to the atomic mass, formulated as follows \cite{pomorski-curv}, 
 \begin{equation}
        \label{20}
        {
                \begin{split}
                        V(a_i)_{\text{LSD}}&= b_{\text{surf}}(1-k_{\text{surf}}I^2)(B_{s}(a_i)-1)\\&+\frac{3}{5}\frac{e^2Z^2}{r_0A^{1/3}}(B_{C}(a_i)-1)\\&+E_{r}^{(0)}(L^2B{j}(a_i)+K^2B_{k}(a_i))\\&+ b_{\text{curv}}(1-k_{\text{curv}}I^2)(B_{\text{curv}}(a_i)-1)\\&+ E_{\text{cong}}^{(0)}(B_{\text{cong}}(a_i)-1)\;.
        \end{split}}
 \end{equation}
 The rotational terms are incorporated following the approaches in refs.\cite{Mazurek2, NADTOCHY}. $E_{\text{r}}^{(0)} = \frac{\hbar^2}{2J_0}$ is the rotational energy coefficient, where $J_0$ is the moment of inertia of the rigid sphere and L and K are the angular momenta with respect to the axis perpendicular and parallel to the symmetry axis, respectively. The deformation of the nucleus is denoted by the parameter $a_i$ where $I=(N-Z)/A$ represents the reduced isospin. $B_s, B_C, B_{\text{curv}}, B_{\text{cong}}, B_j$ and $B_k$ are the deformation-dependent coefficients associated with the surface, Coulombic, curvature, and rotational contributions, respectively, derived using Chebyshev shape parametrization. Additional constant parameters commonly used in LSD models are provided in Table \ref{Tab:1}.\\
 The deformation-dependent terms such as curvature and congruence energy terms play a pivotal role in the fission process. While the surface and curvature terms demonstrate similar behavior at small to moderate deformations. It is observed that the curvature term will dominate significantly as the deformation increases. Understanding this behavior is essential for improving the accuracy of models in predicting the nuclear binding energies and fission barriers. 
 As the nucleus elongates and forms a neck, the congruence energy term ($B_\text{cong}$) dominates, which leads to the fission process. The equation used to determine the congruence energy is, 
 \begin{equation}
        E_{\text{cong}}(a_i) = E_{\text{cong}}^{(0)}(I) B_{\text{cong}}(a_i)
 \end{equation}
 \begin{equation*}
        B_{\text{cong}}(a_i) =
        \begin{cases} 
                2 - \frac{R_{\text{neck}}(a_i)}{R_{\text{frag}}(a_i)}, & \text{for necked-in shapes} \\
                1, & \text{otherwise}
        \end{cases}
 \end{equation*}
 \begin{equation*}
        E_{\text{cong}}^{(0)}(I) = -C_3 \exp(-W_2 |I| C_3)
 \end{equation*}
 where $C_3=10$ MeV and $W_2=42$ MeV. The strength of the neck formation depends on the radius of the neck ($R_{\text{neck}}(a_i)$) and the radius of the nascent light fission fragment ($R_{\text{frag}}(a_i)$), which in turn influences the deformation energy of the fissioning nucleus.  
 \begin{table}
        \caption{Parameters used in LSD model}
        \centering
        \begin{tabular}{l l l}
                \hline\hline
                
                $b_{\text{surf}} = 16.9707$ MeV & \hspace{20pt} $k_{\text{surf}} = 2.2938$   &  \\
                $b_{\text{curv}} = 3.8602$ MeV   & \hspace{20pt} $k_{\text{curv}} = -2.3764$ &  \\
                $r_0 = 1.21725$ fm&   &  \\
                \hline\hline
        \end{tabular}
        \label{Tab:1}
 \end{table}
 
 \subsubsection{Microscopic approach}
 
To complement the macroscopic modelling of the nuclear potential energy, a microscopic description is employed to evaluate the behaviour of individual nucleons in a deformed mean-field potential. This approach provides a more detailed understanding of how the single-particle energy levels evolve with nuclear deformation.
In the present work, the microscopic structure is computed by solving the single-particle Schrödinger equation with a deformation-dependent mean-field Hamiltonian constructed using a finite-range Yukawa folding method \cite{yukawa}. The total single-particle Hamiltonian takes the form,
        \begin{equation}
                H = \frac{p^2}{2m} + V_\text{N} + V_{\text{C}} + V_{\text{S-O}}
        \end{equation}
        The terms represent the kinetic energy, mean-field nuclear potential, Coulomb interaction, and spin-orbit coupling, respectively.
        As the nuclear surface is not sharply defined but exhibits a gradual fall-off in nucleon density, the nuclear, Coulomb, and spin-orbit potentials are all generated from a smooth, diffuse nucleon density distribution. This gradual decrease in density near the nuclear surface is known as surface diffuseness. The density profile, which captures the transition from the dense nuclear interior to the low-density surface region, is through a folding procedure using a finite-range interaction. The folded density is expressed as,     
 \begin{equation}
        \rho(\vec{r}) = \rho_0 \int d^3 r' g(|\vec{r}-\vec{r}'|)
 \end{equation}
 where $\rho_o$ is the constant volume density and $g(|\vec{r}-\vec{r}'|)$ is a Yukawa-folding function with a realistic surface diffuseness. 
 
 \begin{equation}
        |\vec{r}_ - \vec{r}'| = \sqrt{\rho^2 + \rho'^2 - 2 \rho \rho' \cos(\phi - \phi') + (z - z')^2}
 \end{equation}
 
 where, $\rho,$ and $\rho'$ are radial coordinates in cylindrical coordinate system. These coordinates are determined by the Chebyshev shape parametrization as shown in Eq. \ref{1}, which defines the nuclear surface and allows for the exploration of various deformation configurations. Consequently, the resulting $V_\text{N}$, $V_\text{C}$ and $V_\text{{S-O}}$ become functions of the nuclear shape, directly influencing the single-particle energy levels across different deformations.
  \begin{figure*}
        
        \subfigure[]{
                \includegraphics[width=0.49\linewidth]{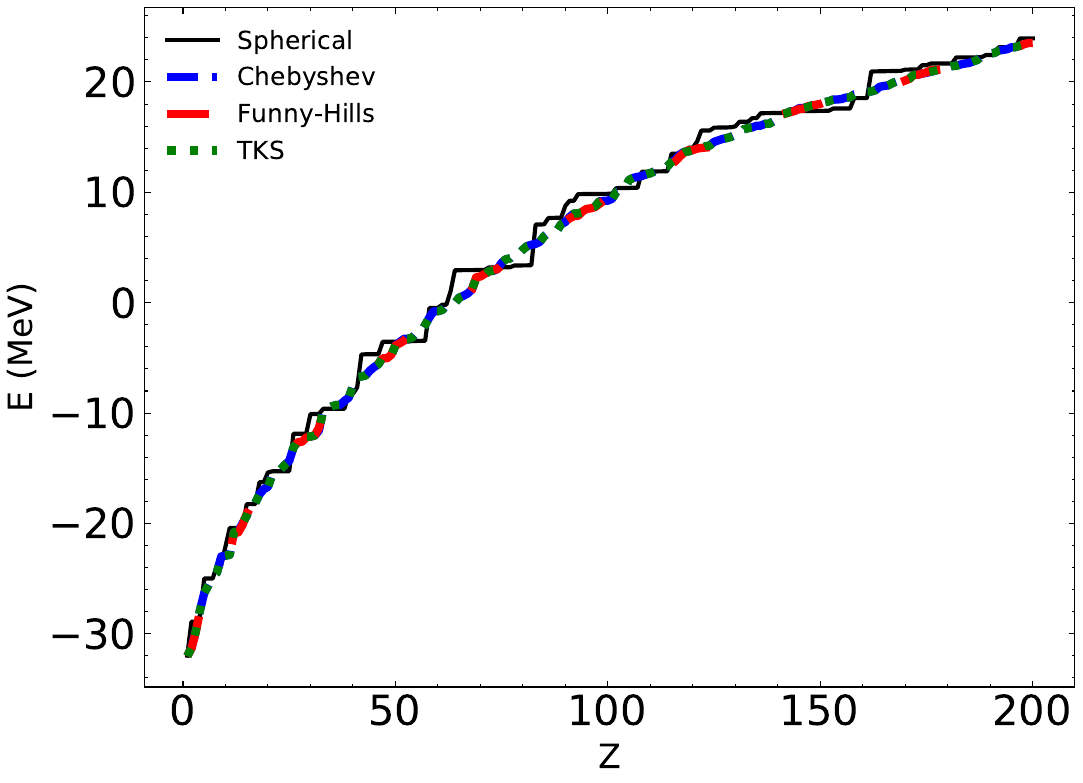}}
        \subfigure[]{
                \includegraphics[width=0.49\linewidth]{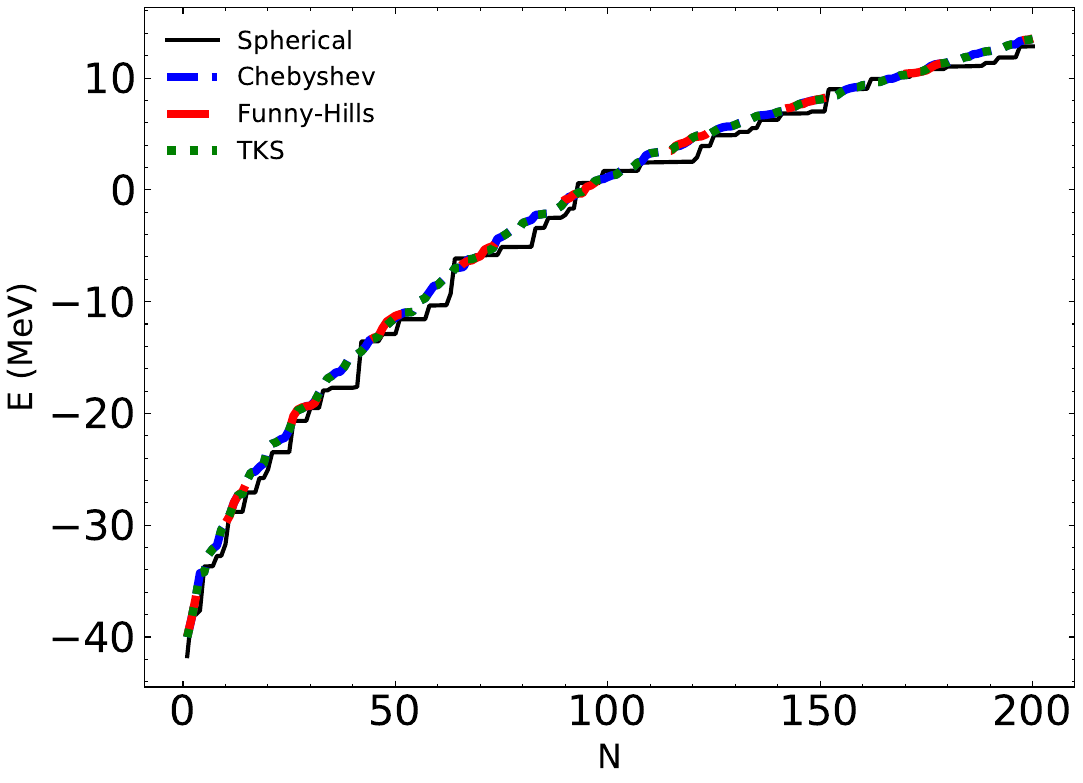}}
        \caption{Single-particle energy levels for (a) protons and (b) neutrons in the nucleus $^{227}$Pa, calculated using different nuclear shape parametrizations: Chebyshev, Funny Hills, TKS. Results for the spherical shape, obtained using the Chebyshev parametrization, are also included for comparison.}
        \label{Fig:7}
 \end{figure*} 
 \section{Results and Discussion}
 \label{sec3}
 In the present study, we focused on the structural changes of  $^{227}$Pa nucleus during the fission process. This is facilitated by explaining the potential energy surface (PES), depicted using the newly introduced Chebyshev shape parametrization within the macroscopic approach based on LSD model. Recently, the different fission modes of the $^{227}$Pa nucleus have been studied, confirming the presence of both symmetric and asymmetric fission modes\cite{Reddy, Saneesh}. In our study, we carefully illustrate the PES for various combinations of deformation parameters, specifically $(a_2,a_3)$, $(a_2,a_4)$ and $(a_2,\eta)$.  While other parameters are fixed at values corresponding to the minimum potential energy of the fissioning nucleus at angular momentum L=0. The deformation parameters $a_2$ and $a_4$ are interconnected, such that $a_2$ primarily influences the elongation and $a_4$ affects the neck formation. The parameters $a_3$ and $\eta$ represents asymmetry and non-axiality in the nuclear shape respectively.
 
 Fig.~\ref{Fig:4}(a) and (b) illustrates the PES within the two-dimensional deformation subspace defined by $( a_2, a_3)$ and $( a_2, a_4)$ respectively. The remaining parameters are fixed at values corresponding to the minimum potential energy of the fissioning nucleus. The probable shape transitions from spherical to an extreme dumbbell shapes in the fission paths are depicted in the upper panel. In Fig.~\ref{Fig:4}(a), the dashed line represents the most probable fission trajectory at $a_3 = 0 $, which implies a symmetric fission path. At  $a_2 = 0$, the nuclear shape is spherical, and further increase in $a_2 $  causes an elongation of the nucleus and transforms to an ellipsoid followed by a symmetric dumbbell configuration at $a_2 = 0.5$. Along this path, the nucleus experiences a single-hump in the potential energy curve. Other minimum energy points corresponding to $a_3$ values from $0$ to $\pm 0.1$, indicates that the left and right asymmetric fission can occur for the $^{227}$Pa nucleus. 
 The Fig.~\ref{Fig:4}(b) reveals a minimum potential energy path to fission starts from  $(0,0)$ and reaches the scission point at $ (0.5, -0.26)$. The scission point spanned over a region of $ a_4 = -0.2$ to $-0.28$ with $a_2 =0.5$ , indicating the possibility of a broader range of  symmetric and asymmetric fission modes. As $a_4$  increases, the curvature and congruence effects influence the total potential energy, which alter the shape and gradually leading to the fission.
 
 Fig.~\ref{Fig:5}(a) and (b) illustrates the PES defined by $( a_2, \eta)$ and $( a_4, a_3)$ respectively. The nuclear shape at the ground state $(a_2=0)$ is deformed to some degree than perfectly spherical. In Fig.~\ref{Fig:5}(b) the first minimum potential energy point is at $(0,0)$ and a minimum energy pathway leading to the fission is observed. This path is characterized by $a_4$ values ranging from $0$ to $-0.25$, with a saddle point at $(-0.08, 0)$. A second potential minimum is identified, extending from $a_3=-0.04$ to $0.04$ for a specific value of $a_4\approx0.24$. This indicates the possibility of multiple fission pathways involving symmetric ($a_3=0$) and asymmetric fission ($a_3\ne0$) modes.
 
 Following the macroscopic approach, we advance to the microscopic description of the nucleus. In this framework, single-particle energy levels are determined by diagonalizing the Yukawa-folded mean-field Hamiltonian within a deformed harmonic oscillator basis. The nuclear shape is characterized using the Chebyshev shape parametrization, and calculations are performed for the nucleus $^{227}$Pa at a specific deformation.
 Fig. \ref{Fig:7} displays the single-particle energy levels for both protons and neutrons in $^{227}$Pa, computed using three distinct shape parametrizations such as Chebyshev, Funny-Hills, and TKS. The deformation parameters are selected to yield identical nuclear shapes, with the following values for chebyshev $a_2=-0.2475,a_3=0,a_4=0$
  and $\eta=-0.15$ ; for Funny-Hills, $c=2.020, h=-0.255, \alpha=0.0,$ and $\eta=-0.15$;  and TKS, $\alpha_2=-0.33, \alpha_3=0,\alpha_4=0$, and $\eta=-0.15$. The equivalence of these shapes is ensured through a transformation relation derived among the parametrizations from Eqs.~\ref{19}, \ref{19-a}.
 The close agreement among the single-particle energy levels obtained from these three parametrizations at the specified deformation validates the robustness of our transformation relation and confirms the applicability of the Chebyshev shape parametrization for microscopic calculations.
 
 \section{Conclusion}
 \label{sec4}
In this study, we have implemented a novel and effective Chebyshev shape parametrization approach to analyze the geometry of atomic nuclei and the energy landscape of nuclear systems within both macroscopic and microscopic frameworks. The deformation parameters derived through volume conservation and center-of-mass constraints have proven essential in describing various structural features of nuclei. In particular, we have elucidated the interconnected roles of $a_2$ and $a_4$ in governing nuclear elongation and neck formation, respectively, while $a_3$ captures asymmetry and $\eta$ accounts for non-axiality.
Our results demonstrate that the Chebyshev shape parametrization provides a more consistent formalism by offering transformation equations that connect it seamlessly with other widely used shape parametrizations. Through the deformation-dependent macroscopic energy coefficients, within the framework of LSD model, we have examined the potential energy surface and explored large amplitude phenomena. Specifically, the PES cross-section of $^{227}$Pa was analyzed to investigate fission pathways and potential shape transitions under various deformations.
Advancing to a microscopic description, we have computed single-particle energy levels by diagonalizing the Yukawa-folded mean-field Hamiltonian in a deformed harmonic oscillator basis, with the nuclear shape described by the Chebyshev parametrization. The results of selected deformations highlight the descriptive power of this approach in characterizing shell structures.
As a future direction, we intend to enhance the potential energy calculations by incorporating shell and pairing corrections within a macroscopic-microscopic framework. This extension will further refine our understanding of nuclear structure and fission dynamics, emphasizing the intricate interplay among deformation parameters and their role in nuclear stability.
\begin{acknowledgements}
K. Jyothish and M. S. Suryan Sivadas acknowledge the financial support provided by the Council of Scientific and Industrial Research (CSIR), India via grant
numbers 09/239(0556)/2020-EMR-1  and  09/0239(11911)/2021-EMR-1 respectively. A. K. Rhine Kumar acknowledges the financial support provided by the Department of Science and Technology (DST), India, (DST/INSPIRE/04/2016/002545) via the DST-INSPIRE Faculty award. This work is also supported by the Science and Engineering Research Board (SERB), India, under Grant Code: CRG/2023/004323.  
   
\end{acknowledgements}

\end{document}